# The possible explanation of neutron lifetime beam anomaly

A. P. Serebrov,* M. E. Chaikovskii, G. N. Klushnikov, O. M. Zherebtsov, A. V. Chechkin.

*Petersburg Nuclear Physics Institute, NRC "Kurchatov Institute",*
*188300 Gatchina, Leningrad District, Russia*

## Abstract

The results of measurements performed using UCN storing method are in good agreement. The latest most accurate measurements of the neutron decay asymmetry and neutron lifetime measurements by storage method are in agreement within the Standard Model. However, there is a significant discrepancy at 3.6σ (1% of decay probability) level with beam method experiment. This article discusses the possible causes of discrepancy in the measurements of the neutron lifetime with beam method experiment. The most probable cause, apparently, is the loss of protons in beam method experiment during storage in a magnetic trap due to charge exchange collisions of protons with the residual gas. The proton becomes neutral and leaves the trap, which leads to a decrease in the number of registered protons, i.e. to a decrease in the probability of neutron decay or to an increase in the measured neutron lifetime.

A. P. Serebrov,* E-mail: serebrov_ap@pnpi.nrcki.ru

## 1. Introduction.

The neutron is the most long-lived of all unstable elementary particles with a lifetime of about 880 seconds. The first experiments to measure the neutron lifetime were carried out more than 70 years ago, yet, even now new researches are carried out and measurement techniques are being improved. Such interest to the topic can be explained by an importance of precise measurements of the neutron lifetime for elementary particle physics and cosmology. However, the relatively high value of the neutron lifetime creates certain difficulties in performing measurements with an accuracy required to validate existing theoretical models. Currently, the accuracy of measurements is of order 0.1%; however, there exists an unresolved problem of a 1% of the neutron lifetime disagreement between results obtained with two main methods.

In Standard Model (SM) of elementary particles the neutron decays into the proton, electron and electron anti-neutrino. There is also a small probability of decay with additional photon or with hydrogen atom in final state:

$n \rightarrow p + e^- + \bar{\nu}_e$      100%
$n \rightarrow p + e^- + \bar{\nu}_e + \gamma$      $(9.2 \pm 0.7) \cdot 10^{-3}$     (1)
$n \rightarrow H + \bar{\nu}_e$      $3.9 \cdot 10^{-6}$

There are two fundamentally different approaches to measure neutron lifetime. The first approach is based on a detection of products of β-decay (protons or electrons) during the period when a neutron beam passes an experimental apparatus. In the second approach, one studies a change in amount of neutrons in an experimental volume with time. Strictly speaking, in these two methods different physical constants are measured: if one detects decay products than probability of exactly the neutron β-decay is measured, while the second method gives an opportunity to use neutron counts to measure total decay probability regardless of a decay channel and final state particles. Even within the SM those are two different probabilities with difference of $4 \cdot 10^{-4}$ % corresponding to the decay with hydrogen atom in final state, but that small fraction of decays is usually neglected. If both methods are considered to measure neutron lifetime that means it is assumed the β-decay is the only possible neutron decay channel.

Registration of decay products is performed in so-called "beam" method of β-decay probability measurements. In this method a beam of cold neutrons passes through a vacuum system in which electrical field is applied to guide decay products to a particle detectors. In later implementations of this method a neutron beam passes through electromagnetic proton trap. Protons from neutrons β-decays are trapped by the field and hence can be stored in trap for a certain time period. After the storing period the protons are guided to the detector (fig. 1). An implementation of the proton trap decreased experimental uncertainties to ≈3s [1].

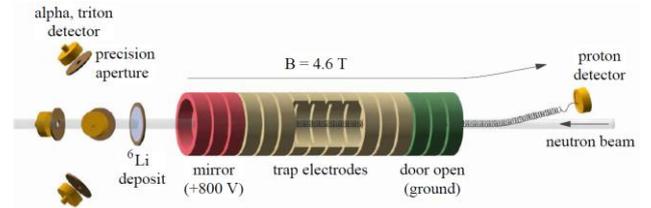

Fig. 1. Experimental apparatus used in the last most precise beam measurement of the neutron lifetime.

To perform experiments using beam method a precise measurements of an initial neutron flux and proton counts are required, because the final result is derived on the basis of the ratio of these two parameters.

The measurements with neutron counting are carried out using storing of ultracold neutrons (UCN). A feature of UCNs is that they can be stored in magnetic or material traps for a long period of time. It is possible due to UCN reflections from walls of material or magnetic traps and a probability to lose a neutron in collision with trap wall can be decreased to very low values. The loss probability in the trap can be decreased to 1-2% of the neutron decay probability, applying the cryogenic material traps [2, 3] and even lower losses are achievable with magnetic traps [4, 5, 6, 7]. That means neutrons can be stored in traps and neutron lifetime can be measured almost directly, introducing small corrections for UCN losses in the trap. The experimental technique is based on filling UCNs into the trap and counting of remaining neutrons after some period. The amount of remaining neutrons is measured for two or more time periods and based on the results a value of the mean neutron lifetime in the trap is calculated.

Due to losses of neutrons in hits with walls the calculated time parameter is less than the free neutron lifetime and it is called a storing time of UCNs in the trap of that specific configuration. Measurements are usually carried out for several geometric configurations of the trap and/or in various energy ranges and the final value of the free neutron lifetime is calculated using an extrapolation to zero loss probability. In experiments with UCN traps one directly measures the total neutron lifetime denoted as $\tau_n$, or total neutron decay probability regardless of the decay products in final state.

The problem of disagreement between two described methods appeared for the first time in 2005, when the result $\tau_n = 878.5 \pm 0.7 \pm 0.3\ s$ [2] was obtained using a cryogenic UCN trap with gravitational locking of neutrons in the trap. For that moment the value of the neutron lifetime according to PDG was $885.7 \pm 0.8\ s$ and the result $\tau_\beta = 886.6 \pm 1.2_{stat} \pm 3.2_{sys}\ s$ that shortly before had been obtained using the beam method [1] was included into derivation of the PDG value, and it confirmed earlier results (see fig. 2). Therefore, the result of the new experiment based on UCN storing deviated from the PDG value for more than 6σ and had significant disagreement with the mentioned beam experiment result. However, further experiments with implementation of magnetic and material traps confirmed the discrepancy [8]. Currently the conventional value of free neutron lifetime according to PDG is $879.4 \pm 0.6\ s$ and it is determined by the results of only UCN storing experiments. Yet results of beam method experiments still have significant disagreement with that value.

In last few years several experiments with UCN storing in magnetic and material traps were carried out. The results are shown in figure 2, the discrepancies between new UCN experiments do not exceed two standard deviations. However, new beam experiments data obtained after appearance of neutron anomaly problem does not exist (the result of beam experiment in 2013 was obtained after additional analysis of data obtained in 2004). And finally, the latest data of measurements of β-decay asymmetry are in agreement with neutron lifetime obtained in experiments with UCN storing [9].

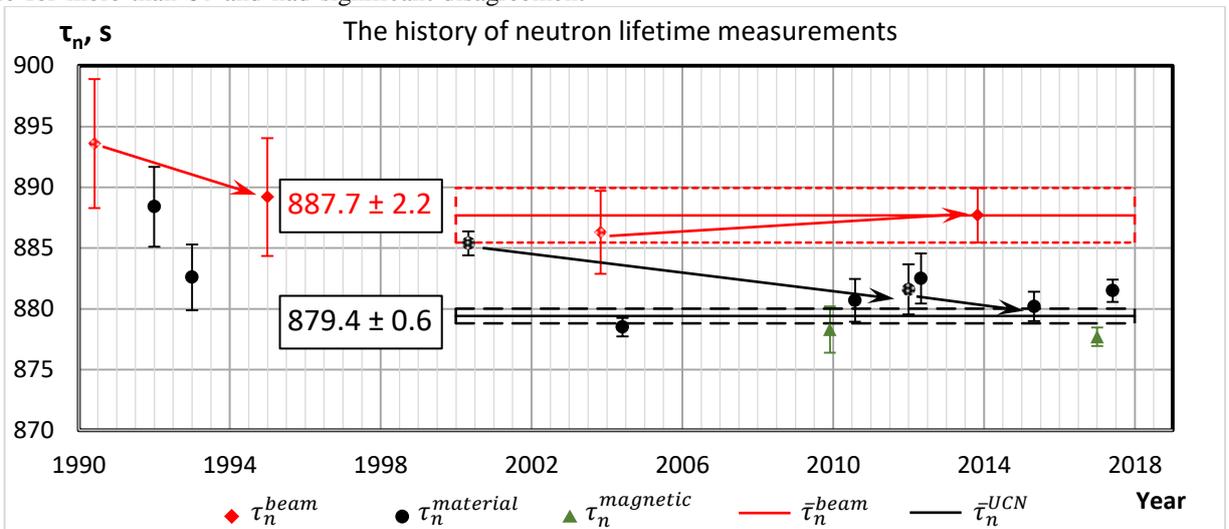

Fig. 2. History of the neutron lifetime measurements since 1990.

For the moment, the reason of this significant disagreement was not discovered. The discrepancy of results of two methods is sometimes called "the neutron anomaly". One can consider two approaches to solve neutron anomaly problem either search for deviations from SM i.e. search for some new physics to explain the discrepancy between the neutron lifetime and β-decay period, or provide analysis of additional systematic errors which were overlooked. Additional errors in UCN storing experiments and various explanations of neutron anomaly with processes beyond SM were considered in ref. [10]. General conclusion of the provided analysis is that explanation of the anomaly lies outside that area.

For the reasons listed above we assume that one should search a solution of neutron anomaly problem in systematic uncertainties of beam method experiments. Detailed analysis of systematic uncertainties in proton counts and measurements of neutron flux were performed in ref. [1]. However, proton losses during storing in magnetic trap and transportation to a detector were considered negligible. Here we want to analyze possible proton losses in the trap.

Recently experimental efforts were focused on UCN storing experiments and no new beam experiments were carried out. Therefore, to solve the neutron anomaly problem it is essential to perform a new more accurate beam experiment. Such experiment is currently under preparation [11], therefore, now it is very important to consider all possible systematic uncertainties, especially those overlooked in previous experiments.

## 2. Residual gas as possible source of systematic error.

Here we want to consider residual gas in vacuum system to be a possible source of systematic errors. For the first time, errors induced by residual gas were considered in ref. [12] and our goal is to extend this analysis and increase its accuracy.

In our work we consider configuration of the experimental setup which was used to obtain the most accurate beam measurements of the neutron lifetime [1]. These measurements rely on accuracy of counting protons emitted in β-decay. The neutron lifetime is calculated based on ratio of proton counts to events in neutron monitor, hence additional proton losses lead to a measured lifetime which is higher than the real value. The proton trap length and hence amount of decays were varied and experimental results were obtained using derivation of an angular coefficient of the curve which describes the dependence of decay number on trap length. Therefore, neglect of proton losses leads to systematic error. Thus an additional term of $(1 - \delta)$ has to be added in the main expression (8) from ref. [1]:

$$\frac{\dot{N}_p}{\dot{N}_{\alpha+t}} = \tau_n^{-1}\left(\frac{\epsilon_p}{\epsilon_0 v_0}\right)(nl + L_{end})(1-\delta) \qquad (2)$$

We present qualitative and quantitative estimations of residual gas effects relying on the experimental setup description from ref. [1]. In this setup a vacuum system consists of a proton trap, proton detector and neutron monitor. Two ion pumps are located on the opposite sides relatively to the trap and they were used to pump out the vacuum system. Importantly, the proton detector and the neutron flux monitor had higher temperature than the proton trap during measurements. Therefore, residual gas pressure was measured in warmer part of the vacuum system. Protons mostly reside in proton trap; therefore the main parameter of our estimations is concentration of residual gas particles in the proton trap which was not directly measured in the experiment.

Essential parameters of residual gas which directly affect proton losses are composition and concentration. Direct measurements of the residual gas pressure were performed by ionization vacuum gauge. Measured electrical current corresponded to pressure of about $10^{-9}$ mbar. This quantity we will consider as a starting point in our analysis. In order to make a model of composition of residual gas we use averaged data of ion pumping [13]. Taking into account concentration and composition in "warm" part of the vacuum system we evaluate concentrations of gases into the trap.

First and foremost, the evaluation requires temperature within the trap. At $10^{-9}$ mbar mean free path of particles far exceeds size of the vacuum system, therefore, molecular trajectories are determined by hits with walls and interaction of gas molecules is negligible. Therefore, particle dynamics depends on vacuum system geometrical configuration and temperature of the walls. The proton trap can be considered as a long vessel with small inlet. Gas molecules, which enter the trap, remain in it for a long time and reach thermodynamic equilibrium with walls.

Consider a following model: a vessel with cold walls and small inlet is located inside closed vessel with warm walls and small gas concentration inside. Initial gas concentration in inner vessel is negligible. Gas from outer vessel enters the inner one and after a few hits with walls cools down to inner vessel temperature, hence gas in inner vessel has the same temperature as the walls. In equilibrium state amount of gas in inner vessel is constant, but warm gas enters it and cold gas leaves. Hence, in equilibrium state warm and cold fluxes at the inner vessel inlet have to be the same: $n_1 v_1 = n_2 v_2$ and $\frac{n_1}{n_2} = \frac{\sqrt{T_2}}{\sqrt{T_1}}$.

That model can be applied to the proton trap and we consider upper bound estimation for concentration of residual gas to be:

$$n = \frac{P}{k\sqrt{T_1 T_2}} \qquad (3)$$

where $P$ is pressure in warm part of the vacuum system, which is measured by ion pump, $k$ is Boltzmann constant, $T_1$ and $T_2$ are temperatures in the warm and cold parts. Warm part we consider to be at room temperature $300K$, and cold part – $4K$ which corresponds to cooling of superconducting solenoid with liquid helium. With $P = 10^{-9}$ mbar we obtain concentration in cold part $n = 2.1 \times 10^8$ particles/cm³. If we consider cold part temperature to be $10K$ or $20K$ we obtain concentrations $1.3 \times 10^8$ particles/cm³ and $n = 9.3 \times 10^7$ particles/cm³.

Besides total concentration, calculations of proton losses also require partial concentrations of gases in the trap. We rely on available data of ion pumping to estimate concentrations in warm part of the trap. Composition of residual gas depends on how the walls were prepared to pumping. High vacuum bake out is especially important because it strongly suppresses amount of water in residual gas. Most surfaces of the vacuum system were baked, but there were few exceptions, for example, the bore of the superconducting solenoid. During the experiment direct measurements of partial pressures of the gases were not performed. For that reason, in our analysis we use data of pumping with various surface preparation technics and obtain a range of possible losses.

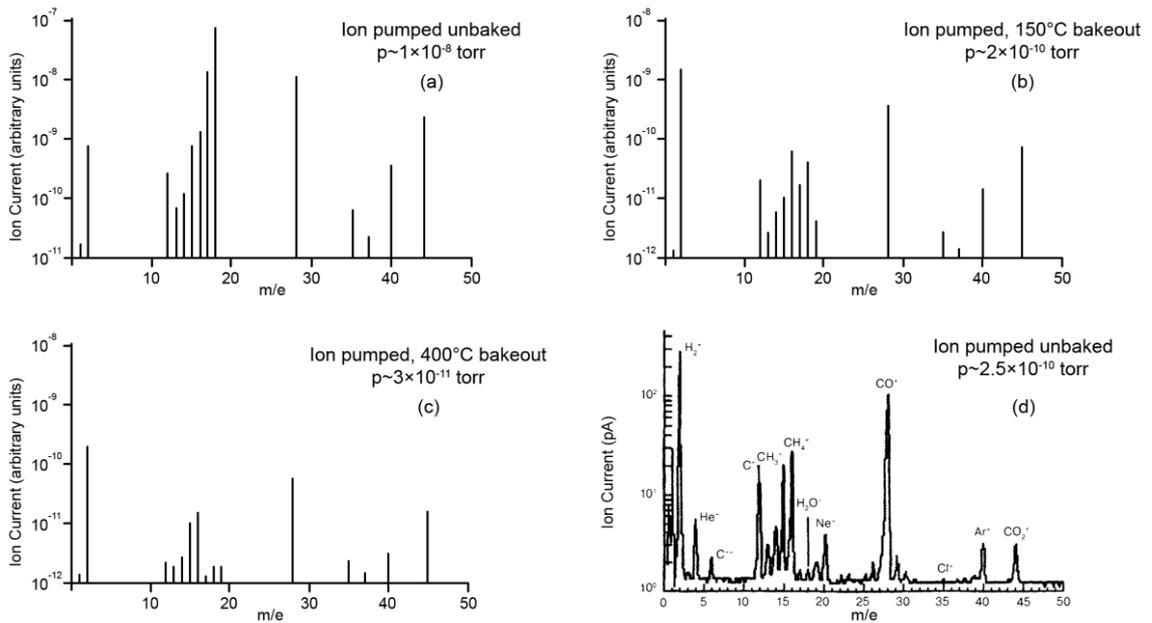

Fig. 3. Mass-spectrum of residual gas after pumping out with ion pump: a –no baking out, b – baking out at 150°C, c –baking out at 400, d – typical spectrum with identified lines.

If a surface was not baked at high vacuum then the main residual gas is water vapor as shown in figure 3.a. If a surface was baked at 150°C then concentration of water significantly decreases and for the most part residual gas consists of hydrogen molecules with addition of $CO$ molecules as shown in figure 3.b. With 400°C bake out water vapor concentration becomes even smaller, but relative parts of $CO$, $CO_2$ and $CH_4$ are increased as shown in figure 3.c.

In all that cases we are interested in relative concentrations, because total concentration is determined by temperature of the proton trap and pressure of residual gas in warm part of the trap.

In figure 3 the data of residual gases concentrations are presented in the form of electrical currents in the measuring system. To extract gas concentrations from the currents one needs to use coefficients which describe effectiveness of registration for each gas and data on fragmentation for each gas, because most molecular gases creates several lines in mass-spectrum. For example, $CO_2$ molecules create signals at $m/e$ equal to 44, 28, 16 and 12 corresponding to ions $CO_2^+$, $CO^+$, $O^+$, $C^+$. The data required to extract concentrations from the currents were taken in ref. [13]. Table 1 shows relative concentrations of gases which have biggest parts in residual gas.

**Table 1.** Composition of residual gas in vacuum system pumped out by ion pump with various surface preparation technics.

| Conditions\Gas | $H_2$ | $CH_4$ | $H_2O$ | $CO$ | $Ar$ | $CO_2$ |
|---|---|---|---|---|---|---|
| Unbaked $p \sim 10^{-8}$ torr | 0.4 | 1.5 | 88.7 | 6.5 | 0.4 | 2.2 |
| Bake out at 150 °C $p \sim 2 \cdot 10^{-10}$ torr | 63.0 | 4.6 | 4.7 | 18.9 | 1.5 | 5.6 |
| Bake out at 400 °C $p \sim 3 \cdot 10^{-11}$ torr | 52.5 | 11.8 | 1.5 | 22.3 | 1.6 | 7.4 |

The obtained spectra can be used to approximate composition of residual gas in the proton trap in our calculations. This approximation can be used if adsorption in the trap can be neglected. At the very least, it can be used to obtain an upper limit of proton losses.

Using concentration and composition of residual gas one can perform qualitative and quantitative estimation of how it affects the proton losses. Two processes can contribute into the proton losses: elastic scattering and charge exchange interaction of protons with residual gas molecules.

### 3. The proton losses in elastic scattering.

Electrostatic system of the experiment consists of cylindrical cavity with solenoid, which creates static magnetic field, and a system of electrodes. The electrode system can be divided in two parts 1) "trap electrodes" or grounded electrodes with zero potential, 2) "mirror" electrodes – the electrodes situated at the ends of the trap and which determine an area where axial component of the velocity $v_z$ changes its sign (see fig. 1).

There are three modes of the trap operation: 1) trapping protons, 2) counting proton, 3) cleaning the trap. β-decay protons created in the trap move between mirror electrodes in trapping mode, because maximal kinetic energy of the decay protons is less than potential barrier between grounding and mirror electrodes. In counting mode electric field is applied in the trap to guide protons to a detector. Finally, in cleaning mode electric field push all charged particles out of the trap and prepare it to next measurement cycle. Further we consider proton dynamics in details to estimate possible losses in trapping mode.

In the general case, a proton created in the trap has complicated movement limited in space (see Fig. 4.) with oscillations of three types: 1) axial – along the axis of the solenoid which is caused by gap in potentials of electrodes, 2) azimuthal – rotation around the solenoid axis, 3) cyclotron – rotation around guiding center. Guiding center moves along magnetic field lines and drifts in electric field with velocity $v_E = (E \times B)/B^2$ orthogonal to magnetic field lines. Each oscillation has its own period.

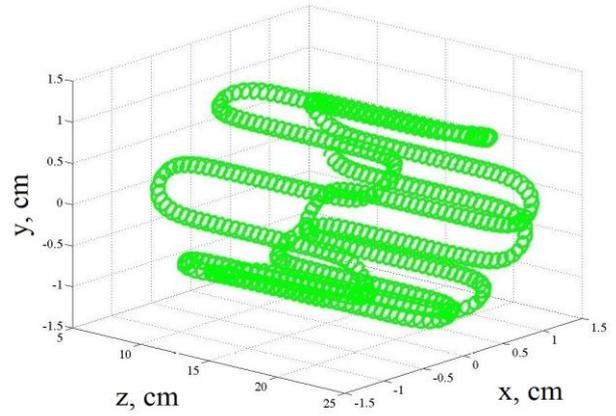

Fig. 4. Trajectory of a proton with initial kinetic energy 700eV in the trap. Oz – solenoid axis.

In the experimental setup trap radius, magnetic field strength and neutron beam width are tuned in a way that β-decay protons always reach the detector without hitting the wall of the trap. However, losses caused by elastic scattering were not taken into account. To estimate these losses we carried out special simulation of proton motion within the trap considering elastic scattering at residual gas. In simulation we used elastic scattering cross-sections of protons at hydrogen, nitrogen and oxygen atoms.

In simulation we assumed that elastic scattering at an atom occurs at some point of proton trajectory and as a result proton velocity changes its direction with scattering angle $\theta$. Using differential cross-section of elastic scattering and energy conservation law for each pre-hit proton velocity $v_0$ probability distribution function of scattering angles $\theta$ was calculated. The next step was to calculate how post-hit proton velocity depends on its pre-hit velocity and scattering angle, and using obtained post-hit velocity we calculate a new mean time between hits $\tau$. Calculated dependencies were represented in form of tables and these tables were used in the simulation of proton trajectories.

From the simulation results one can conclude that if there is high concentration of residual gas in the system and if initial radial coordinate is close to inner radius of the electrodes 13mm, after 5-7 scatterings in angles close to 90 degrees the proton trajectory can cross the cylindrical surface. In most scenarios losses occur in end regions because magnetic field lines bend there and cyclotron radius increases.

In fig. 5 one can see how a projection of a proton trajectory on a plane orthogonal to solenoid axis changes with increasing of residual gas concentration. At low concentrations the trajectory is projected on a set of points located between two concentric circles and hence changes of radial coordinate are limited. Starting from certain concentration which depends on target atom type frequency of hits increases to values at which protons can be lost due to unbalanced fluctuations of radial coordinate (see fig. 5.b). Further increase of the concentration results in more curved proton trajectories which are not limited by cylindrical surfaces. From the results obtained in the simulations one can conclude that taking into account parameters of the considered experiment only low pressure scenario takes place, while scenarios b and c require much higher concentration of the residual gas.

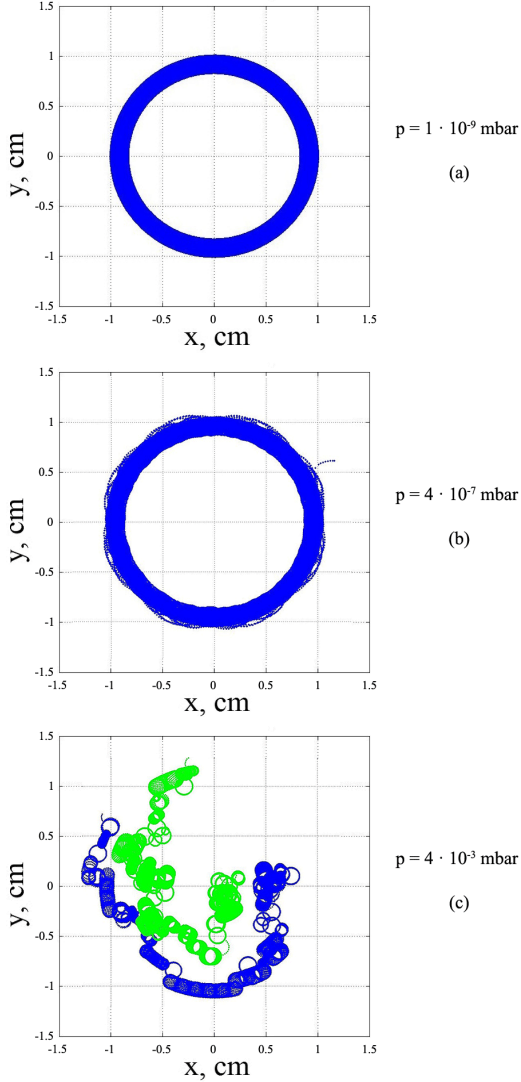

Fig. 5. Projections of proton trajectories on a plane orthogonal to solenoid axis obtained in simulations of elastic scattering in proton trap for various concentrations of Oxygen atoms:
a) small concentration, b) medium concentration,
c) high concentration.

In summary, a proton in the trap can be lost due to elastic scattering only after several scatterings in angles close to 90°, taking into account that mean storing time is 5 ms. In case of oxygen atoms with pressure equal to total calculated pressure in the considered experimental setup the proton losses are only $8.3 \cdot 10^{-6}$ that is much less than neutron anomaly and currently available experimental accuracy of neutron lifetime measurements. Our simulations with oxygen atoms reveal that to have 1% losses residual gas concentration have to be at least 3 orders of magnitude higher than the calculated value. From these results one can conclude that at currently available experimental accuracy of neutron lifetime measurements in setups with proton trapped pumped out to ultra-high vacuum the losses caused by elastic scattering at residual gas are negligible.

## 4. Charge exchange process as losses source

Now we consider charge exchange process between protons and residual gas atoms: $H^+ + A \rightarrow H + A^+$. Charge exchange cross-section of protons with velocities much less than classical electron velocity in atom has rather big value for many gases and significantly exceeds elastic scattering cross-section, hence this process causes systematic error. β-decay protons do not get additional acceleration in proton trap so their energy is less than 750 eV and velocity is more than an order of magnitude less than classical velocity of an electron in an atom.

Using data about partial pressures in residual gas after pumping by ion pump we can make a list of atoms and molecules which interact with protons in the trap. For each gas we calculate its specific proton loss coefficient using expression:

$$K_i = \int \sigma(E) v(E) t_m P(E) dE \qquad (4)$$

where $\sigma(E)$ – charge exchange cross section of this gas, $v(E)$ – proton velocity at given energy, $t_m$ – mean time of proton storing in the trap, $P(E)$ – normalized β-decay protons spectrum. Coefficient $K_i$ has dimension $cm^3$.

In calculations we used proton charge exchange cross-sections available from various published sources. The list of considered gases corresponding loss coefficients and sources of data about cross-sections are presented in table 2. The corresponding charge exchange cross-sections are shown in figure 6. Since there are no helium peaks in mass-spectra on figure 3 and charge exchange cross section for the helium atoms is four orders of magnitude smaller than of atomic hydrogen it is omitted from figure 6 and the tables.

**Table 2.** Loss coefficients of protons for gases having largest parts in residual gas.

| Target | Loss coefficient $cm^3$ | Data source |
|---|---|---|
| $H_2$ | $2.426 \cdot 10^{-11}$ | [14] |
| $CH_4$ | $3.332 \cdot 10^{-10}$ | [15] |
| $H_2O$ | $3.050 \cdot 10^{-10}$ | [16] |
| $CO$ | $1.896 \cdot 10^{-10}$ | [16, 17] |
| $Ar$ | $4.872 \cdot 10^{-11}$ | [18] |
| $CO_2$ | $2.617 \cdot 10^{-10}$ | [16] |

In general, the accuracy of measured cross-sections is about 10-20% depending on gas. For most of the gases one can find analytical approximating expression which fits data with accuracy of the same level as experimental accuracy. These expressions were used in calculations. For other gases we used interpolation of existed experimental data. In energy region below $100\ eV$ for some gases experimental data do not exist nor has higher uncertainties, but, fortunately, that region contributes very little to the calculated coefficients. Only 8% of the protons have energies below $100\ eV$ and such protons have shorter paths in the trap and hence low probability of losses due to charge exchange process.

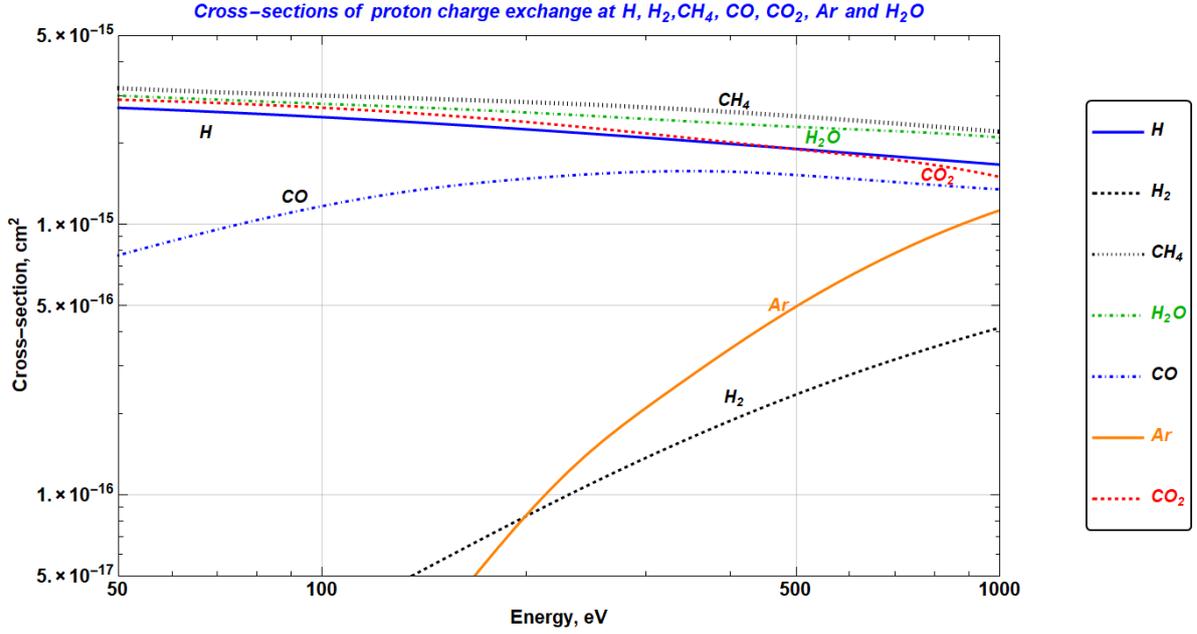

Fig. 6. Charge exchange cross sections of molecules which have largest parts in residual gas.

Charge exchange process results in creation of neutral hydrogen atom which leaves the magnetic proton trap and a slow molecular ion of the gas at which the process occurred. This ion is trapped and continues its motion at the trajectory analogous to proton trajectory. It is obvious that charge exchange in case of atomic hydrogen doesn't affect the losses because it generates the same proton with lower energy and it can be accelerated by the electric field and can be registered by the detector as well. In our analysis we also don't consider further interactions of molecular ions created in the charge exchange process with the residual gas molecules since probability of such interactions are proportional to the loss coefficient squared from table 2. After storing period ions (and charged molecules) along with protons are accelerated in electric field to $30\ keV$ and reach the detector. A charge exchange process results in signal loss if an ion would be absorbed or back scattered in detectors dead layer or if the signal of the ion is under the detector discriminator threshold. Therefore, passing of molecular ions through the detector dead layer should be analysed.

Initial molecular ion after acceleration has total energy about $30\ keV$ and to mimic proton signal it has to pass the dead layer of proton detector and leave enough energy in sensitive layer to form signal above pulse-height discriminator. Therefore, for each gas there is a probability that charge exchange process actually results in the loss of the signal. In order to calculate that probabilities we carried out simulation using program SRIM-2013 [19]. Figure 7 represents the stopping ranges for the hydrogen and oxygen ions along with corresponding ionization within the detector.

We simulated passing of molecular ions through the detector dead layer and sensitive part of the detector. The main goal was to estimate the energy that goes into ionization of the sensitive layer and comparison of that energy with the threshold.

In our simulation we assume that molecular ion breaks into atoms in collision with the detector and atoms passes through the detector separately but simultaneously. The kinetic energy of the initial molecule spreads between atoms proportionally to atomic masses. For each simulated molecule we calculate energy of ionization of the sensitive detector layer and simulate its signal in the detector. The parameters of the simulation were tuned in a way that simulation of proton signal coincide with experimental results obtained in ref. [1].

As a result of the simulation we obtained probabilities to loss proton signal after charge exchange process for each considered gas. We call them loss probabilities and denote as $P_i$. The results of the simulations are shown in fig. 8 and corresponding loss probabilities are listed in table 3.

**Table 3.** Loss probabilities of molecules having largest parts in residual gas.

| Molecule | Loss probability $L_i$ |
|---|---|
| $H_2$ | 0 |
| $CH_4$ | 0.20 |
| $H_2O$ | 0.26 |
| $CO$ | 0.60 |
| $Ar$ | 0.77 |
| $CO_2$ | 0.95 |

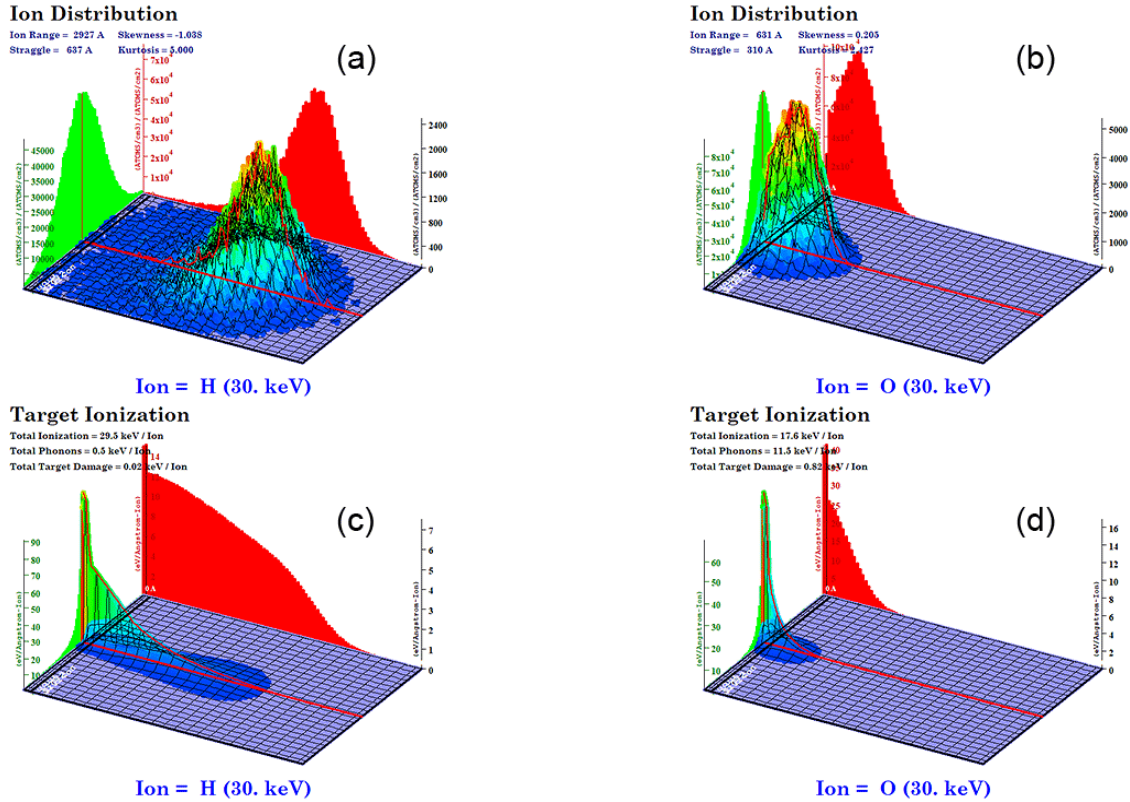

Fig. 7. Ion distribution and target ionization plots for hydrogen (a) and (c) and oxygen (b) and (d) with initial energy of 30 $keV$ passing through a 100Å layer of gold than 25Å layer of silicon oxide and finally stopping in a pure silicon.

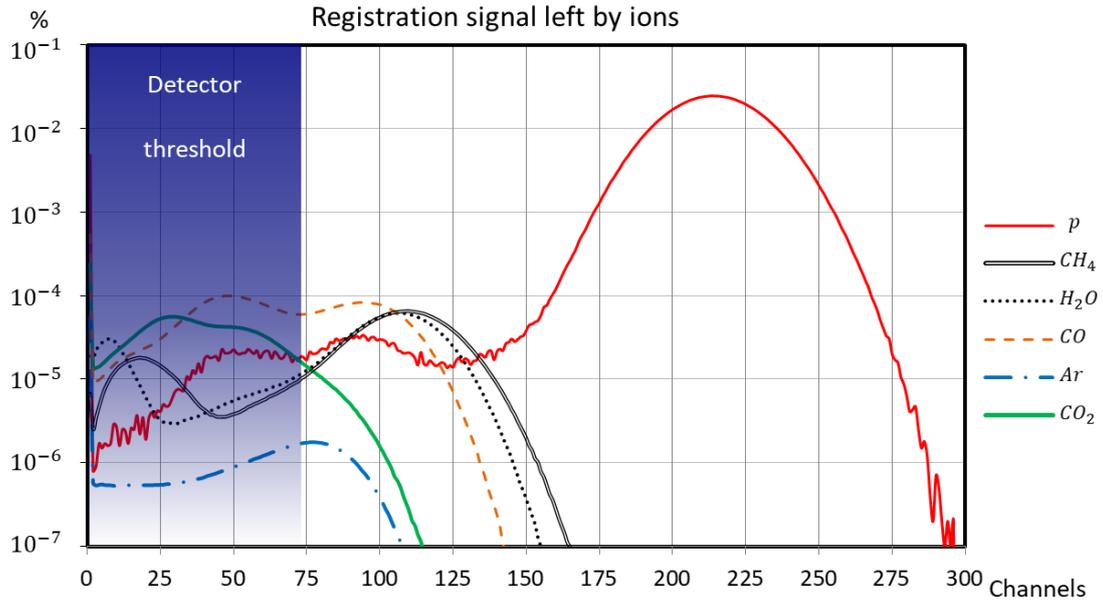

Fig. 8. Simulation of signals of molecular ions for molecules having largest part in residual gas.

The fraction of protons that is lost due to charge exchange process at a gas is a product of the concentration, loss coefficient, and loss probability of that gas. To obtain total proton losses we should sum losses over all gases:

$$\delta_{loss} = \sum_i n_i \cdot L_i \cdot K_i \quad (5)$$

Using equation (5) coefficients from tables 2 and 3 and gas concentrations from table 1 we calculated losses for 5 temperatures inside proton trap for three mass-spectra of atoms and molecules in residual gas corresponding to three conditions of vacuum system surface baking. In all scenarios we assumed that pressure of residual gas in vicinity of ion pump is $10^{-9}$ mbar, the value presented in ref. [1]. That assumption was made to normalize the results of the analysis in order to reveal the effects caused by variation of residual gas composition. It is well-known that different baking conditions lead to different limits for residual gas pressure in vacuum system. In particular, vacuum with pressure $10^{-9}$ mbar is almost unachievable without baking, while with baking at $400°C$ the pressure can be made significantly lower than $10^{-9}$ mbar. We will return to it in further analysis. The results with nominal vacuum $10^{-9}$ mbar and gas composition corresponding to conditions of baking from table 1 are listed in table 4.

**Table 4.** Proton losses (correction of the neutron lifetime) with nominal vacuum $10^{-9}$ $mbar$.

| Surface preparation temperature / Effective of residual gas in the trap | No baking $10^{-9}$ mbar | **Baking at 150 °C $10^{-9}$ mbar** | Baking at 400 °C $10^{-9}$ mbar |
|---|---|---|---|
| 4K | 1.8% (-15 s) | **0.89% (-7.8 s)** | 1.1% (-9.8 s) |
| 10K | 1.1% (-9.8 s) | **0.56% (-4.9 s)** | 0.70% (-6.2 s) |
| 20K | 0.79% (-6.9 s) | **0.40% (-3.5 s)** | 0.50% (-4.4 s) |
| 100K | 0.35% (-3.1 s) | **0.18% (-1.6 s)** | 0.22% (-1.9 s) |
| 300K | 0.20% (-1.8 s) | **0.10% (-0.9 s)** | 0.13% (-1.1 s) |

From these results one can conclude that baking of the surface of a vacuum system can significantly decrease proton losses mostly because it reduces water fraction in residual gas. It is important to notice that the simulation was performed in assumption that the layer of gold had surface density of $20$ $\mu g/cm^2$ which is the minimum value of all used in the experiment. The probability coefficients increase with increasing of dead layer thickness and difference between baking condition become more intense.

In the next step we take into account that different surface preparation conditions lead to different limits on vacuums. Limits on vacuums are shown in figure 2: $1 \cdot 10^{-8}$ torr in case of no baking out, $2 \cdot 10^{-10}$ $torr$ in case of baking out at $150°C$ and $3 \cdot 10^{-11}$ $torr$ in case of baking out at $400°C$. Proton losses and neutron lifetime corrections according to limit vacuum conditions are shown in table 5.

**Table 5.** Proton losses (correction of the neutron lifetime) in vacuum corresponding to limit achievable with the baking.

| Surface preparation temperature / Effective of residual gas in the trap | No baking $1 \cdot 10^{-8}$ $torr$ | Baking at 150°C $2 \cdot 10^{-10}$ $torr$ | Baking at 400 °C $3 \cdot 10^{-11}$ $torr$ |
|---|---|---|---|
| 4K | 23% (-206 s) | 0.24% (-2.1 s) | 0.04% (-0.39 s) |
| 10K | 15% (-130 s) | 0.15% (-1.3 s) | 0.03% (-0.25 s) |
| 20K | 10% (-92 s) | 0.11% (-0.9 s) | 0.02% (-0.17 s) |
| 100K | 4.7% (-41 s) | 0.05% (-0.4 s) | 0.01% (-0.08 s) |
| 300K | 2.7% (-23 s) | 0.03% (-0.2 s) | 0.005% (-0.05 s) |

It is rather obvious, that first results presented in first column of table 5 have little to do with considered experimental setup, because in experiment the vacuum measured near the ion pump was an order of magnitude better and the setup was baked. But the first column demonstrates the importance of baking of the vacuum system surfaces because it removes water which significantly affects results.

Second column from table 5 do not exactly satisfy the experimental conditions because measured residual gas pressure was 5 times higher. The closest to experimental conditions is the results from the second column of table 4 and it is indicated with bold text. The limit pressure in vacuum of $2 \cdot 10^{-10}$ $torr$ was not achieved because some parts of the vacuum system could not be baked. Superconducting solenoid and semiconductor detector of protons are most important of them.

Finally, the results presented in third column lead to conclusion that to achieve accuracy better than $1$ $s$ with cold trap vacuum of about $3 \cdot 10^{-11}$ $torr$ is required. Meanwhile, achieving $1$ $s$ accuracy with warm trap requires vacuum $2 \cdot 10^{-10}$ $torr$, which can be reached with moderate baking at temperature $150°C$. A warm trap have to be separated from a cold superconducting solenoid which has temperature $4K$.

Finally, we summarize our analysis of possible systematic errors in the experiment [1]. The results presented in table 6 can be considered as closest estimation of the experiment [1] despite the fact that not all surfaces were baked out.

**Table 6.** Correction of the neutron lifetime in case of vacuum $10^{-9}$ $mbar$ which was achieved in the considered experiment in vicinity of ion pump at room temperature. Composition of residual gas corresponds to baking of all vacuum system at 150 °C.

| Surface preparation / Effective temperature of residual gas in the trap | **Baking at 150°C $10^{-9}$ $mbar$** |
|---|---|
| 4K | **-7.8 s** |
| 10K | **-4.9 s** |
| 20K | **-3.5 s** |
| 100K | **-1.6 s** |
| 300K | **-0.9 s** |

As shown in table 2, the gases, which are in most part responsible for proton losses in the trap, are $H_2$, $CH_4$, $CO$, $CO_2$ and $H_2O$. Concentration of water significantly decreases after baking and it also can be frozen out at low temperatures. $H_2$, $CH_4$, $CO$, $CO_2$ have vapor pressure $1 \cdot 10^{-8}$ $torr$ at temperatures 4K, 32K, 27K and 76K correspondingly [20]. These gases also condensate at surfaces at higher temperatures, but above the surfaces with listed temperatures the pressure would be $1 \cdot 10^{-8}$ $torr$ which is even an order of magnitude higher than the pressure used in calculations.

## 5. Conclusions

The most probable source of systematic error in experiment [1] is residual gas in proton trap. Large cross-section of charge exchange process of slow protons at residual gas molecules results in significant systematic error even in case of high vacuum.

Cooling of the trap does not decrease systematic error, after detailed analysis we conclude that cooling in fact increases that error. Therefore, to increase accuracy of measurements of neutron lifetime with beam method one should use proton trap with walls at room temperature.

Loss coefficients of the gases must be taken into account in constructing vacuum system and its surfaces. The direct analysis of residual gas composition is mandatory for neutron lifetime measurements with beam method.

Water has big loss coefficient and it is the main cause of systematic errors if the surfaces were not baked out in high vacuum. Therefore design of the experimental setup has to permit warming up of all surfaces to at least 150°C.